\documentclass[
reprint,
superscriptaddress,
preprintnumbers,
 amsmath,amssymb,
 aps,
]{revtex4-1}

\usepackage{graphicx}
\usepackage{comment}
\usepackage{dcolumn}
\usepackage{bm}
\usepackage{xcolor}
\definecolor{ao}{rgb}{0.0,0.5,0.0}

\begin{document}


\title{On the question of dynamic domains and critical scattering in cubic methylammonium lead triiodide}

\author{Nicholas J. Weadock}
 \affiliation{SSRL Materials Science Division, SLAC National Accelerator Laboratory, Menlo Park, CA 94025}
\author{Peter M. Gehring}%
\affiliation{NIST Center for Neutron Research, National Institute of Standards and Technology, Gaithersburg, MD 20899}%
\author{Aryeh Gold-Parker}
 \affiliation{Department of Chemistry, Stanford University, Stanford, CA 94305
}%
\author{Ian C. Smith}
\affiliation{Department of Chemistry, Stanford University, Stanford, CA 94305
}%
\author{Hemamala I. Karunadasa}
\affiliation{Department of Chemistry, Stanford University, Stanford, CA 94305
}%
\affiliation{Stanford Institute for Materials and Energy Sciences, SLAC National Accelerator Laboratory, Menlo Park, California 94025}
\author{Michael F. Toney}
\email{mftoney@slac.stanford.edu}
\affiliation{SSRL Materials Science Division, SLAC National Accelerator Laboratory, Menlo Park, CA 94025}

\date{\today}

\begin{abstract}
We investigate the hypothesis of dynamic tetragonal domains occuring in cubic CH$_3$NH$_3$PbI$_3$ using high-resolution neutron spectroscopy to study a fully deuterated single crystal. The \textit{R}-point scattering above the 327.5 K tetragonal-cubic phase transition is always resolution-limited in energy and therefore inconsistent with dynamic domain predictions. This behavior is instead consistent with the central peak phenomenon observed in other perovskites. The scattering may originate from small, static, tetragonal-phase domains nucleating about crystal defects, and the temperature dependence demonstrates the transition is first-order.  
\end{abstract}

\maketitle


Hybrid organic-inorganic metal-halide perovskites (HOIPs) are a novel class of semiconductor with unusual and promising optoelectronic properties for high-efficiency devices. The HOIP family exhibits significant dynamical disorder due to weak bonding, and for hybrid variants, the additional rotational degrees of freedom of the organic cations.\cite{leguy_dynamics_2015, chen_rotational_2015} This disorder is believed to affect optoelectronic and other properties,\cite{gold-parker_acoustic_2018, beecher_direct_2016, quarti_structural_2016, whalley_perspective_2017} although the connection is not well understood. The dynamical disorder also correlates with phase transitions in the prototypical hybrid HOIP, CH$_3$NH$_3$PbI$_3$ (herein MAPI), which  transitions between orthorhombic, tetragonal, and cubic phases at 162 K and 327.5 K, respectively.\cite{weller_complete_2015} The cubic-tetragonal transition in MAPI proceeds as a condensation of the soft, transverse acoustic, cubic \textit{R}-point phonon.\cite{beecher_direct_2016, comin_lattice_2016} The cubic \textit{R}-point becomes the $\Gamma$-point of the tetragonal phase, giving rise to elastic Bragg scattering.\cite{beecher_direct_2016, comin_lattice_2016, whitfield_structures_2016} X-ray inelastic scattering studies have observed \textit{R}-point scattering in the cubic phase of MAPI, leading to reports of tetragonal domains that are both small and potentially dynamic.\cite{beecher_direct_2016, whalley_perspective_2017, comin_lattice_2016} Computational results at 650 K (above the experimentally observed decomposition temperature \cite{stoumpos_semiconducting_2013}) indicate sub-ps lifetimes for these domains.\cite{quarti_structural_2016} In addition, the hypothesis of dynamic tetragonal domains has been used to explain the lack of variation in the optoelectronic properties of MAPI-based devices between the two phases.\cite{whitfield_structures_2016}

Materials that exhibit soft-mode-driven structural phase transitions often display diffuse ``critical scattering" arising from structural fluctuations near the critical temperature $T_C$, such as the postulated dynamical domains.\cite{stirling_critical_1996} Apparent critical behavior has been studied extensively with neutron and X-ray scattering methods in perovskite compounds including SrTiO$_3$, KMnF$_3$, RbCaF$_3$, CsPbCl$_3$, CsPbBr$_3$, and MAPbCl$_3$.\cite{hirota_neutron-_1995, shapiro_critical_1972, shirane_q_1993, nicholls_determination_1987, ryan_observation_1986, gibaud_comparison_1990, ravy_$mathrmsrtio_3$_2007, holt_dynamic_2007, fujii_neutron-scattering_1974, songvilay_lifetime-shortened_2018, songvilay_common_2019} 

In these materials, neutron inelastic scattering experiments have revealed certain soft-mode transitions that exhibit an anomalous central peak (CP), centered at $\hbar \omega = 0$. First reported in SrTiO$_3$ in 1971 by Riste \textit{et al}., \cite{riste_critical_1971} the evidence to date indicates that the CP is resolution-limited in energy and therefore a static (infinite lifetime) phenomenon; it thus differs from critical scattering, which is dynamic in nature. Several theories of the CP origin have been proposed.\cite{cowley_are_1996} Shapiro and co-workers demonstrated that anharmonic renormalization of the soft-mode phonon results in a scattering function consisting of phonon sidebands plus a central component centered at zero energy.\cite{shapiro_critical_1972} Other studies have proposed that the CP arises from domains of the low-temperature structure nucleating at defects.\cite{ryan_observation_1986, hastings_central-peak_1978, gibaud_critical_1987} A second, longer length scale has been observed in high-resolution X-ray and neutron scattering studies of critical scattering; however, this narrow component is unrelated to the CP because it is absent in the bulk and is due to static Bragg scattering from regions of the low-temperature phase nucleating near the sample surface.\cite{mcmorrow_length_1990, hirota_neutron-_1995, neumann_origin_1995, gehring_absence_1995, holt_dynamic_2007, ravy_$mathrmsrtio_3$_2007, hunnefeld_influence_2002}

We use high-resolution neutron spectroscopy, with an energy resolution at least 10 times better than that of current X-ray inelastic scattering techniques,\cite{beecher_direct_2016, comin_lattice_2016} to characterize the cubic-phase \textit{R}-point scattering in MAPI. Data obtained from constant-\textbf{Q} scans (E-scans), where the energy transfer ($\hbar \omega$) is varied at a fixed momentum transfer \textbf{Q}, and elastic \textbf{Q}-scans, where \textbf{Q} is varied at fixed $\hbar \omega = 0$, were compared to models representing the dynamic-domain and CP hypotheses. The models are assessed in terms of lineshape goodness-of-fit and temperature dependence of peak intensity and linewidth. We find that the \textit{R}-point scattering is resolution-limited in energy but not in \textbf{Q}, and thus consistent with a manifestation of the CP phenomenon and not dynamic tetragonal domains. The weak temperature dependence and wave-vector (momentum) dependence of the \textit{R}-point scattering is compared to theoretical predictions of the CP origin and differs distinctly from that reported in inorganic oxide- and fluoride-based perovskites.

Scattering at the cubic \textit{R}-point \textbf{Q} = $\frac{1}{2}(1\,3\,3)$ for a fully deuterated single crystal of CD$_3$ND$_3$PbI$_3$ (d$_6$-MAPI) was characterized using the cold (SPINS) and thermal (BT4) neutron triple-axis spectrometers at the NIST Center for Neutron Research (NCNR). The d$_6$-MAPI crystal was oriented in the $(1\,0\,0)\perp(0\,1\,1)$ scattering plane as described previously.\cite{gold-parker_acoustic_2018} A fully deuterated sample was used to reduce the background from the large incoherent scattering cross-section of hydrogen. The MA$^+$ cation rotates nearly freely in the cubic phase, resulting in a quasielastic scattering (QES) contribution to the total \textit{R}-point scattering.\cite{chen_rotational_2015, leguy_dynamics_2015} Quasielastic scattering of this nature produces a signal that varies slowly with \textbf{Q}, even with a fully deuterated sample, as deuterium, nitrogen, and iodine all exhibit small incoherent scattering cross-sections of 2, 0.5, and 0.3 barns, respectively. Therefore, we subtract these QES contributions to the total \textit{R}-point scattering as explained below.

Four different energy resolutions were obtained by varying the neutron final energy $E_f$ and adjusting horizontal beam collimations. The elastic incoherent instrumental energy resolution half-width at half-maximum (HWHM) in each case was measured with a vanadium standard on SPINS and calculated for BT4 with the ResLibCal software.\cite{farhi_ifit:_2014} The instrumental wave-vector resolution (HWHM) was determined from Gaussian fits to elastic \textbf{Q}-scans measured along both [100] (longitudinal) and [011] (transverse) directions relative to the cubic phase $(2\,0\,0)$ and $(0\,1\,1)$ Bragg peaks. The spectra were reduced and analyzed with the NCNR Data Analysis and Visualization Environment (DAVE) software package.\cite{azuah_dave_2009} All fitted parameters are reported with errors of $\pm$ one standard deviation. Further details are provided in the Supplementary Materials section.

Figure \ref{fig:fig1}a) shows E-scans measured at \textbf{Q} = $\frac{1}{2}(1\,3\,3)$ and 340 K ($T_C + 12.5$ K) for all experimental configurations, scaled to the same peak intensity. The energy linewidths are resolution-limited in each case. By contrast, the wave-vector linewidths of elastic \textbf{Q}-scans measured along [100] and [011] are not resolution-limited (see Fig. S5) and show little dependence on the instrumental \textbf{Q}-resolution. 

\begin{figure}
    \centering
    \includegraphics{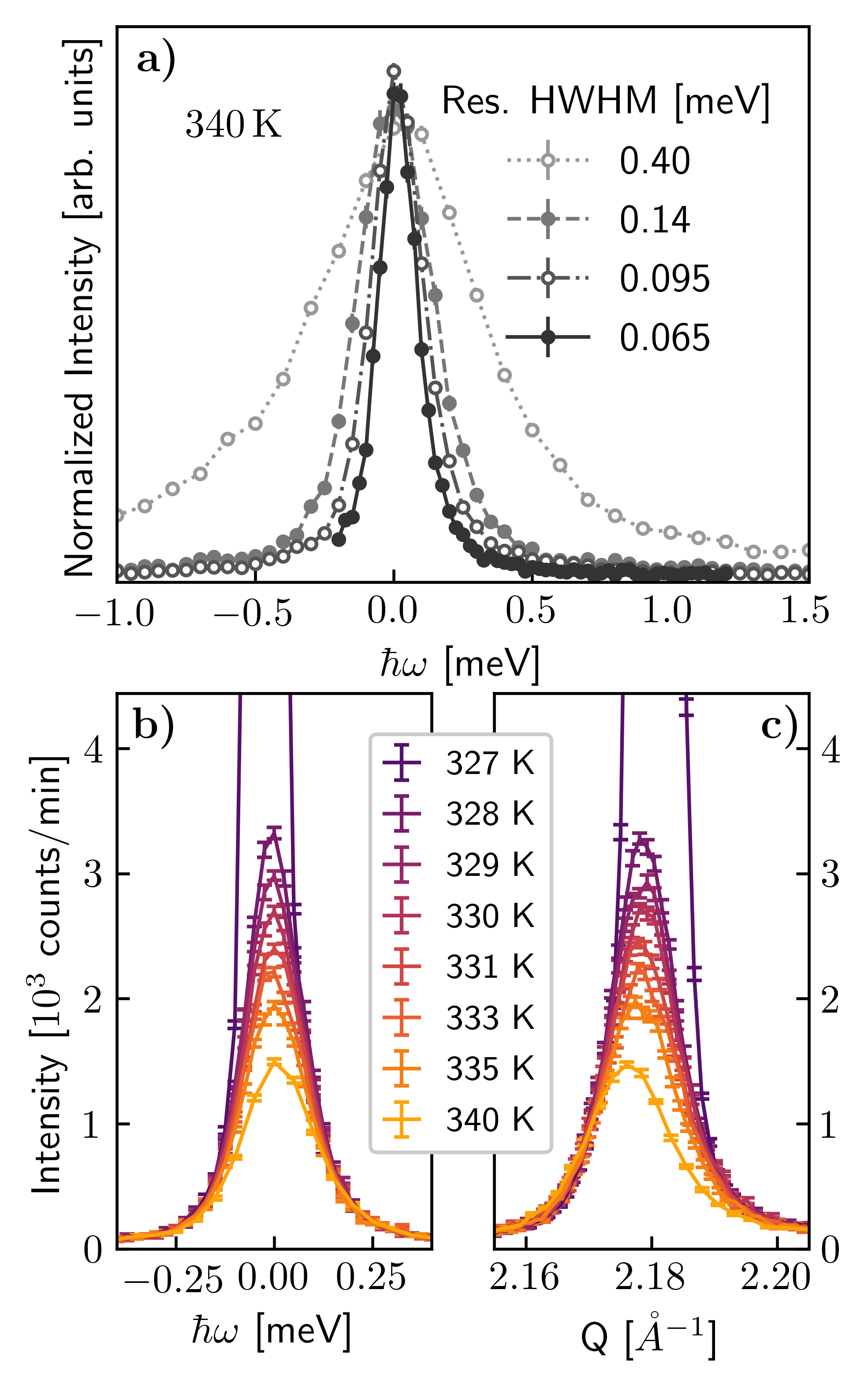}
    \caption{Scattering at the cubic \textit{R}-point \textbf{Q} = $\frac{1}{2}(1\,3\,3)$ in d$_6$-MAPI measured with thermal and cold neutrons. In a), the width of the scattering matches the energy resolution. Data were taken at 340 K on BT4 (dotted line) and SPINS (3 other spectra). \textit{R}-point scattering b) energy dependence and c) wave-vector dependence (at $\hbar\omega = 0$) as a function of temperature measured with an energy resolution HWHM of 0.095 meV ($E_f = 4$ meV).}
    \label{fig:fig1}
\end{figure}

To characterize the temperature dependence of the \textit{R}-point scattering, E-scans and elastic \textbf{Q}-scans were measured at \textbf{Q} = $\frac{1}{2}(1\, 3\, 3)$ on cooling from 340 K – 327 K with an energy resolution of 0.095 meV ($E_f = 4$ meV). The best resolution of 0.065 meV HWHM ($E_f = 3.5$ meV) has reduced dynamic range and cannot cover the full \textit{R}-point scattering profile. Figures \ref{fig:fig1}b) and \ref{fig:fig1}c) display E-scans and \textbf{Q}-scans measured along [100] at the indicated temperatures on SPINS. No phonon peaks are observed at any temperature in the \textit{R}-point E-scans up to $\hbar\omega = 4$ meV. This is not unexpected, as extremely short phonon lifetimes are indistinguishable from background at the zone boundary.\cite{gold-parker_acoustic_2018} The scattering intensity increases slowly on cooling to 328 K but jumps at 327 K, as the crystal undergoes a phase transformation. The cubic-tetragonal transition temperature $T_C$ for d$_6$-MAPI is identified to be $327.5\,\pm0.5$K, consistent with previous reports.\cite{poglitsch_dynamic_1987}

\begin{figure}
    \centering
    \includegraphics{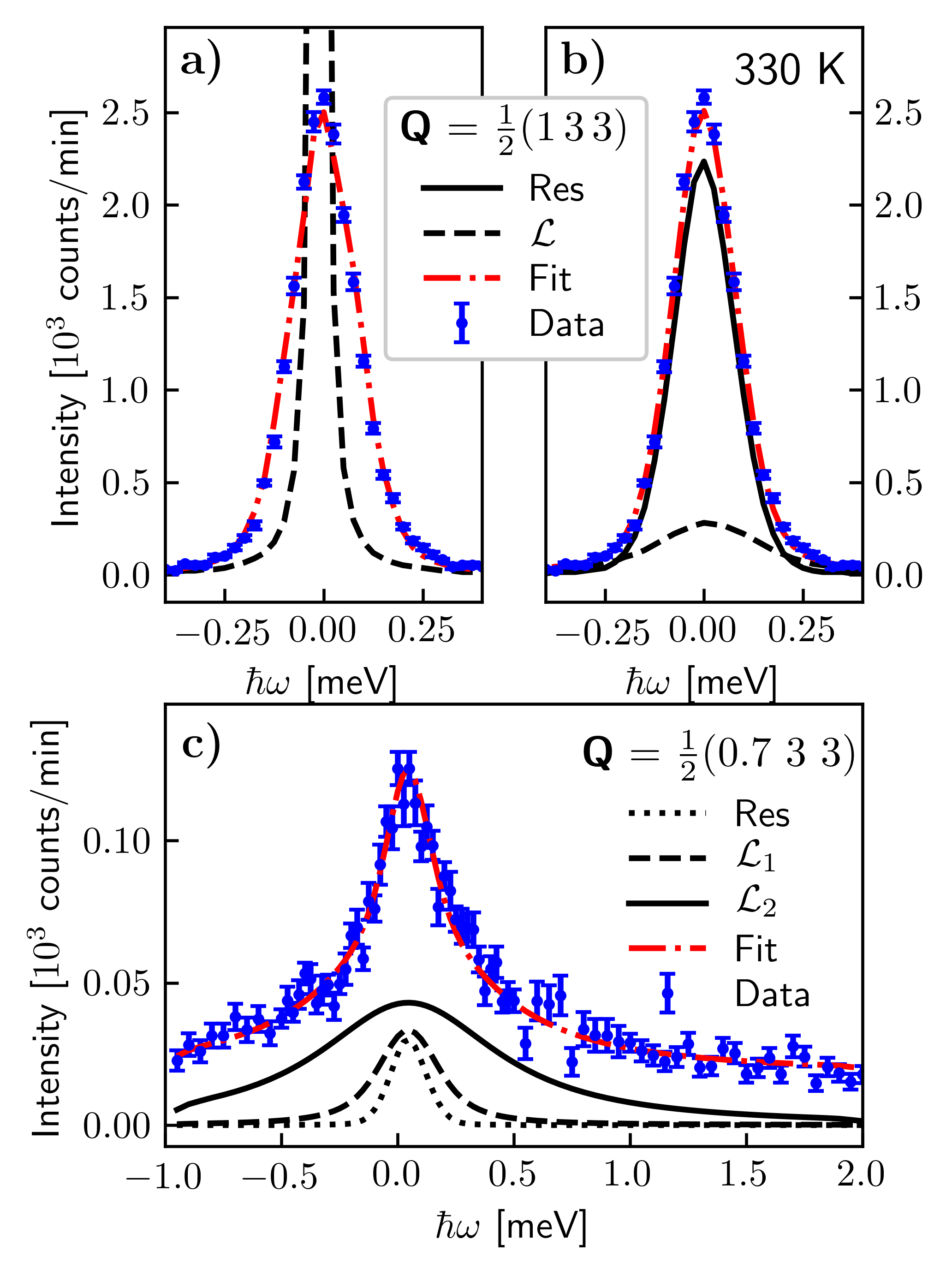}
    \caption{Fits to the QES-subtracted \textit{R}-point E-scan (resolution HWHM = 0.095 meV) at 330 K using the a) dynamic-domain (reduced-$\chi^2=3.41$) and b) CP models (reduced-$\chi^2=1.42$). The resolution-broadened total fits are plotted as a red dash-dot line, the Lorentzian components $\mathcal{L}$ as black dashed lines, and the resolution-broadened delta function as a solid black line. The dynamic-domain model does not fit the tails nor peak as well as the CP model. c) Fit (dash-dot red line) of the QES offset from the R-point to a model consisting of two $\mathcal{L}$ (dashed and solid black lines) and a delta function (dotted black line) convolved with the instrument resolution function, plus a flat background (not shown).}
    \label{fig:fig2}
\end{figure}

To obtain the QES contributions to the \textit{R}-point scattering, we measured the scattering at \textbf{Q} = $\frac{1}{2}(0.7\,\, 3\,\,3)$ slightly offset from the \textit{R}-point at 330 K, as shown in Fig. \ref{fig:fig2}c). The QES was modeled as a delta function to represent the elastic incoherent scattering plus two Lorentzian functions corresponding to fast methyl- and ammonium- group rotations about the C-N bond axis and slower molecular reorientations perpendicular to the C-N axis. Fitted linewidths yield mean residence times of 1.5 ps and 6.8 ps for fast and slow reorientations, respectively, within the range of reported values.\cite{chen_rotational_2015, leguy_dynamics_2015} The QES scattering from \textbf{Q} = $\frac{1}{2}(0.7\,\, 3\,\,3)$ at 330 K was then subtracted from \textit{R}-point scattering at all temperatures.

The dynamic-domain and CP hypotheses are tested by comparing results from fitting these models to the QES-subtracted \textit{R}-point E-scans. The dynamic-domain model consists of a single Lorentzian $\mathcal{L}$, whereas the CP model is the sum of a $\delta$-function (representing  energy-resolution-limited scattering) and a $\mathcal{L}$ to account for remaining QES or other effects. Both models are convolved with the resolution function during the fit. Representative fits of the dynamic-domain and CP model to 330 K spectra are provided in Fig.~\ref{fig:fig2}a) and b). At all temperatures below 340 K, the reduced-$\chi^2$ statistic for the CP model is less than that of the dynamic-domain model indicating the CP model better describes the data. A comparison of goodness-of-fit is provided in the Supplemental Materials.

\begin{figure}
    \centering
    \includegraphics{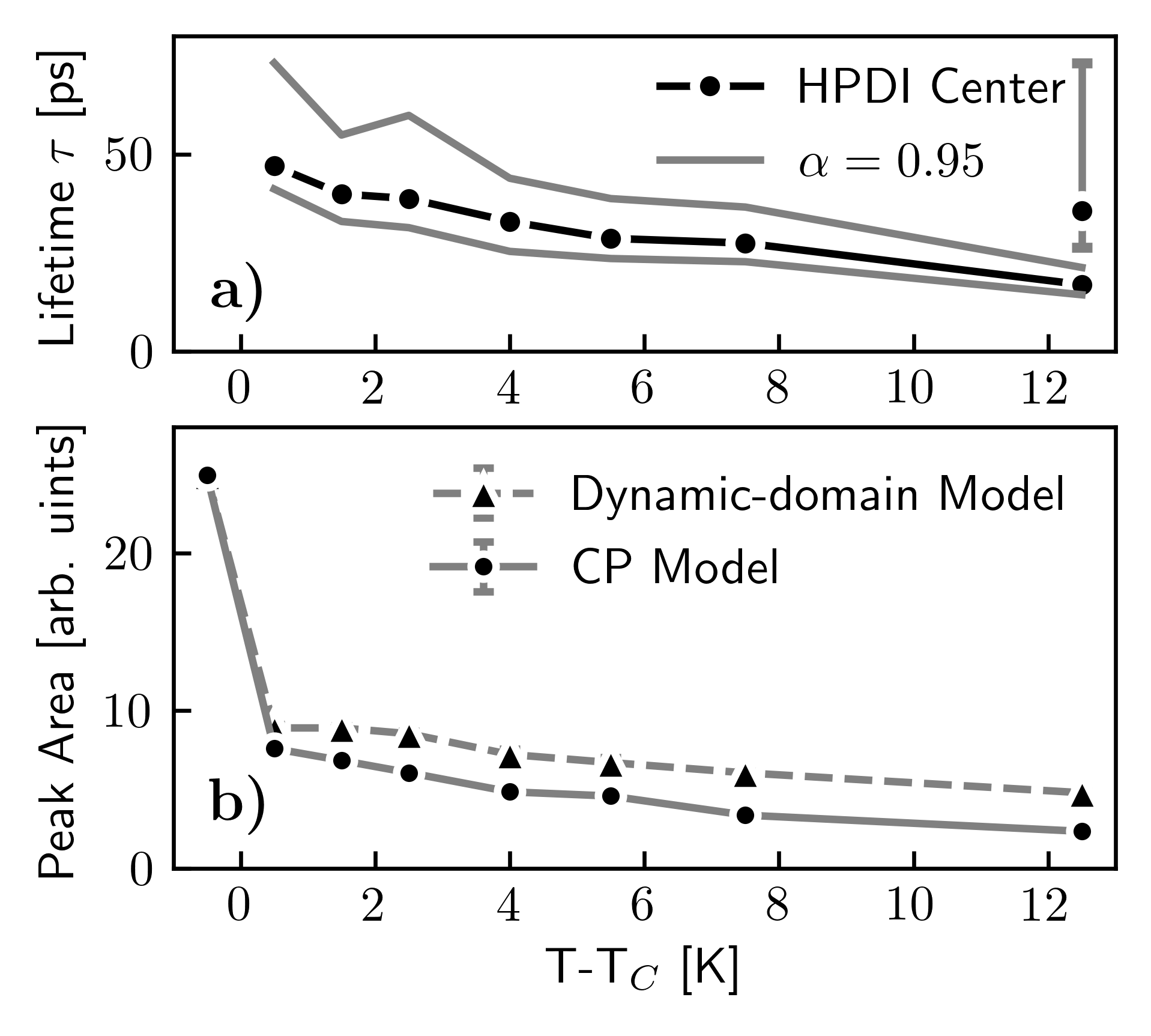}
    \caption{a) Minimum $\tau$ (and 95\% confidence interval $\alpha$) of tetragonal domains determined from maximum likelihood estimation. Temperature-dependent $\tau$ are determined from \textit{R}-point E-scans with resolution HWHM = 0.095 meV. The single point at $T - T_C = 12.5$ K corresponds to an \textit{R}-point E-scan with resolution HWHM = 0.065 meV. b) Peak area obtained from \textit{R}-point E-scans, the CP model corresponds to the resolution-broadened $\delta$-component only. Error bars may be smaller than the corresponding marker.}
    \label{fig:fig3}
\end{figure}

Within the dynamic-domain hypothesis, the lifetime $\tau$ of dynamic tetragonal domains is calculated from the HWHM of the $\mathcal{L}$ as $\tau = \hbar/\mathrm{HWHM}$.\cite{maradudin_scattering_1962} We determine a minimum $\tau$ for dynamic domains by constructing a highest posterior density interval (HPDI) from the log likelihood that the \textit{R}-point E-scans are described by a resolution-broadened $\mathcal{L}$ of varying HWHMs. Details are provided in the Supplemental Materials. Minimum $\tau$ range from $\tau=16.9$ ps to $47.0$ ps at 340 K and 328 K, respectively. These results are plotted in Fig. \ref{fig:fig3}a) and show an unexpectedly weak dependence on temperature. The same analysis was performed for 340 K data with resolution HWHM = 0.065 meV, with a corresponding minimum $\tau=35.6$ ps. These $\tau$ are an order of magnitude longer than both molecular rotation and zone-boundary phonon lifetimes.\cite{chen_rotational_2015, leguy_dynamics_2015, gold-parker_acoustic_2018}

Figure \ref{fig:fig2}b) demonstrates that under the CP model the scattering intensity is almost entirely described by the resolution function at 330 K. Figure \ref{fig:fig3}b) plots the integrated intensity of the resolution-broadened $\delta$ component through $T_C$. Temperature-dependent area fractions of the $\delta$ and $\mathcal{L}$ components are plotted in Fig. S7. The linewidths of the $\mathcal{L}$ component are converted to $\tau$, which slowly decreases with temperature (Fig. S8). 

\begin{figure}
    \centering
    \includegraphics{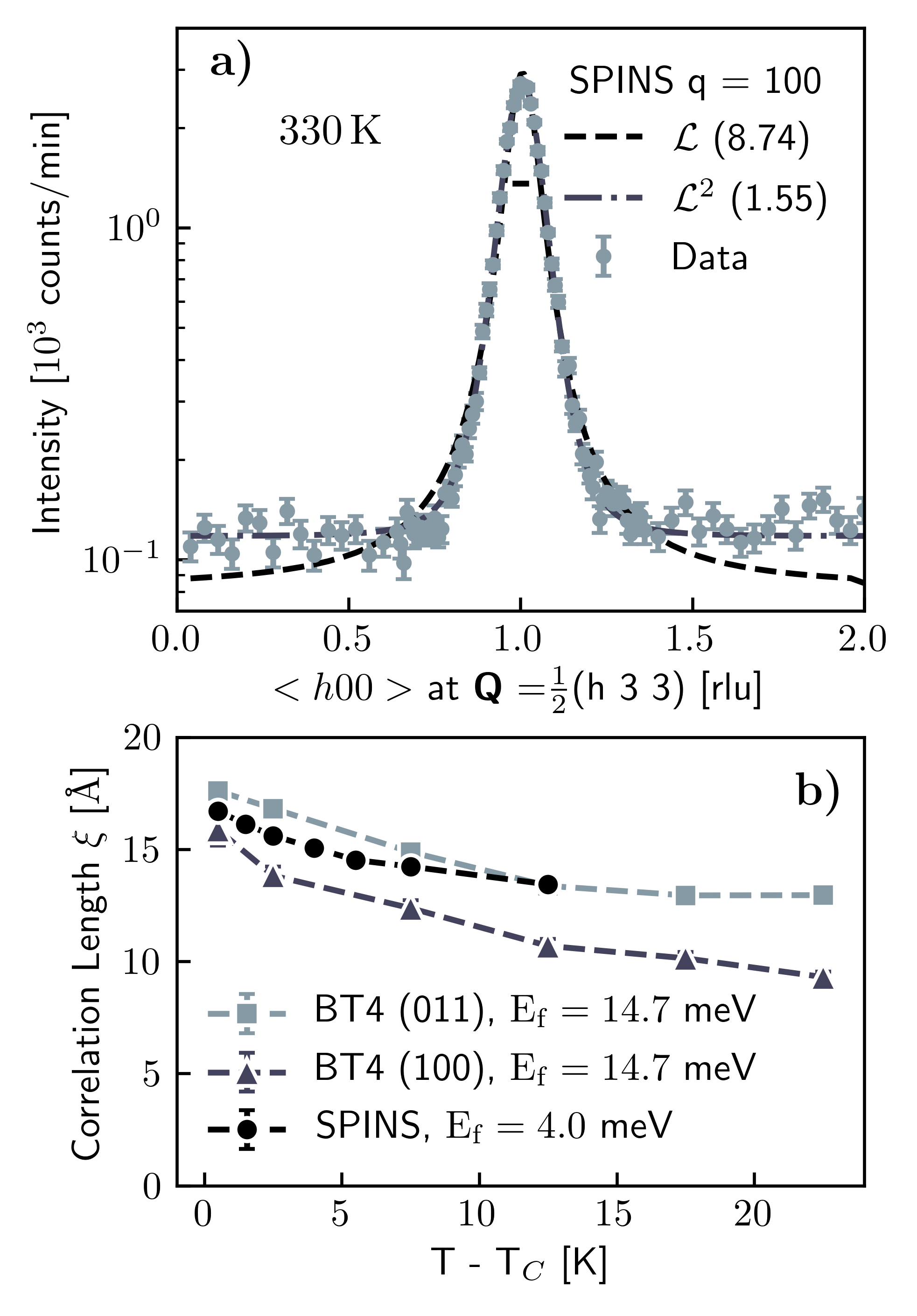}
    \caption{a) Representative elastic \textbf{Q}-scan from SPINS at 330 K along [100] at the \textit{R}-point. The data (gray markers) are fit to both $\mathcal{L}$ (dashed) and $\mathcal{L}^2$ (dash-dot) lineshapes, with corresponding reduced-$\chi^2$ values indicated in the legend. Resolution full-width at half-maximum is indicated with a black line. b) Correlation lengths $\xi$ extracted from $\mathcal{L}^2$ linewidths for \textit{R}-point scans along [100] (SPINS, BT4) and [011].}
    \label{fig:fig4}
\end{figure}

To gain further insight into the two hypotheses, we analyze the momentum dependence of the \textit{R}-point scattering. Elastic \textbf{Q}-scans centered on $\frac{1}{2}(1\,\,3\,\,3)$ were measured along [100] on SPINS and along [100] and [011] on BT4. \textbf{Q}-scans of critical scattering are expected to narrow towards the resolution limit as $T\,\rightarrow\,T_C$, whereas the CP is broader than the \textbf{Q}-resolution HWHM at all temperatures.\cite{ryan_observation_1986, shirane_q_1993} The data at each temperature were fit with either a $\mathcal{L}$ or $\mathcal{L}^2$ function, convolved with the instrumental wave-vector resolution function, plus a flat background, as proposed by CP models.\cite{shapiro_critical_1972, shirane_q_1993, halperin_defects_1976, gibaud_critical_1987-1} The best lineshape was determined by comparing reduced-$\chi^2$ statistics. All scans measured on SPINS are best described by a $\mathcal{L}^2$ lineshape. Figure \ref{fig:fig4}a) plots representative fits for SPINS spectra at 330 K; the $\mathcal{L}^2$ lineshape is a significantly better fit than the $\mathcal{L}$. Scans measured on BT4 (not shown) fit equally well to either lineshape. 

The linewidths extracted from these $\mathcal{L}^2$ fits to elastic \textbf{Q}-scans were converted to correlation lengths $\xi$ using the relation $\xi = (\mathrm{HWHM}/0.64)^{-1}$ \cite{shirane_q_1993} and are plotted in Fig. \ref{fig:fig4}b). Correlation lengths range from 10 - 18 $\mathrm{\AA}$ just above $T_C$ with no significant anisotropy in \textbf{Q}. These $\xi$ are quite small, nearly an order of magnitude less than those measured in RbCaF$_3$ and SrTiO$_3$ at $(T-T_C) = 1\,\mathrm{K}$ and 10 K, respectively.\cite{hirota_neutron-_1995, shirane_q_1993, ryan_observation_1986} The observed temperature dependence is very weak; any critical exponents extracted from inverse power law fits would range from 0.07 – 0.17.

We now compare the two hypotheses in terms of the fitted parameters. The measured HWHM for all experimental configurations in Fig. \ref{fig:fig1}a) match the corresponding resolution HWHM, indicating that the \textit{R}-point scattering is resolution-limited in energy and thus consistent with the CP phenomenon. It could be argued that this observation is the result of a sufficiently narrow $\mathcal{L}$ convolved with a broad enough resolution function to appear resolution-limited.\cite{stirling_critical_1996} In this case the lifetime of tetragonal domains is at least $\tau~=~35.6$ ps (Fig.~\ref{fig:fig3}a), however, all other aspects of the \textit{R}-point scattering are not consistent with this picture. First, the lifetime, peak area, and correlation length (Figs.~\ref{fig:fig3}a,b; Fig.~\ref{fig:fig4}b) do not diverge near $T_C$, showing only weak temperature dependencies. Second, although the scattering is resolution-limited in energy, it is not resolution-limited in \textbf{Q}. This behavior is hard to reconcile with critical scattering from dynamic-domains but is fully consistent with the neutron CP phenomenon.\cite{shapiro_critical_1972, shirane_q_1993} Finally, the goodness-of-fit for the CP model is the better than the dynamic-domain model at all temperatures below 340 K.

Our analysis indicates that the \textit{R}-point scattering observed in d$_6$-MAPI is a manifestation of the CP phenomenon. The remaining $\mathcal{L}$ component of the CP model is unlikely due to overdamped phonons or critical scattering because the linewidth is too narrow and the intensity and $\tau$ decrease on cooling to $T_C$ (Figs. S7, S8).\cite{gold-parker_acoustic_2018} Given that it is a relatively small fraction of the total scattering, we believe it is more likely the result of an imperfect QES subtraction.  

Resolution-limited scattering has been observed at the \textit{M}- and \textit{X}-points near the phase transitions in MAPbCl$_3$ and CsPbBr$_3$, potentially indicating the existence of the CP phenomenon in these materials as well.\cite{songvilay_lifetime-shortened_2018, songvilay_common_2019} By contrast, critical scattering with a dynamic component has been observed at the \textit{R}-point in CsPbBr$_3$ near the tetragonal-orthorhombic transition.\cite{songvilay_common_2019} No analysis of the origin of this dynamic component was provided. 

In the formalism of phonon renormalization as the origin of the central peak, the lineshape of the elastic \textbf{Q}-scan is demonstrated to be dependent on the renormalized mode frequency $\omega_0$, quasi-harmonic soft-mode frequency $\omega_{\infty}$, and a renormalization coupling constant $\delta^2 = \omega^2_{\infty} - \omega^2_0$.\cite{shapiro_critical_1972, shirane_q_1993} When $\omega^2_{\infty} \gg \delta^2$, the central peak is described by a $\mathcal{L}^2$ lineshape; when $\omega^2_{\infty} \approx \delta^2$  the lineshape is $\mathcal{L}$. As the phonons are so heavily damped near the zone boundary, we cannot measure $\omega_{\infty}$. However, we find that a $\mathcal{L}^2$ lineshape best describes all elastic \textbf{Q}-scans, and this suggests that $\omega^2_0 > 0$ meV at all temperatures. This implies a first-order because the renormalized frequency decreases towards $\omega^2_0 = 0$ for a second-order transition, as seen in SrTiO$_3$.\cite{occhialini_negative_2018,scott_soft-mode_1974} In fact, as shown in Fig. S12, the lattice parameter exhibits a discontinuous jump at $T_C$, thus confirming the first-order character of the cubic-tetragonal transition in d$_6$-MAPI.

A $\mathcal{L}^2$ lineshape may arise from scattering by domains of the low-temperature phase nucleating around isolated defects. \cite{halperin_defects_1976, gibaud_critical_1987-1, ryan_observation_1986, imry_influence_1979} Imry and Wortis demonstrated that the presence of these defect-induced domains result in a smearing of a first-order phase transition about a $T_C$.\cite{ryan_observation_1986, gibaud_critical_1987-1, imry_influence_1979} The $\mathcal{L}^2$ lineshape observed with SPINS in Fig. \ref{fig:fig4}a) could indicate that the CP results from static tetragonal domains, $10-18~\mathrm{\AA}$ in diameter (2-3 unit cells), nucleating around defects within the cubic matrix. This result, however, differs significantly from other perovskites, as scattering of this type is only observed from surface domains one to two orders of magnitude larger than the $\xi$ in Fig. \ref{fig:fig4}b). \cite{hirota_neutron-_1995, shirane_q_1993, ryan_observation_1986, mcmorrow_length_1990, gibaud_critical_1987-1, beecher_direct_2016} A smaller $\xi$ could indicate weaker coupling between lead iodide octahedra, consistent with observations of extremely short acoustic phonon mean free paths near Brillouin zone boundaries and the fact that the lattice constant in MAPI (6.30 \AA) is much larger than that in conventional inorganic perovskites like SrTiO3 (3.90 \AA).\cite{gold-parker_acoustic_2018}

We note that the CP area and $\xi$ do not diverge as $T \rightarrow T_C$ as is observed in SrTiO$_3$, KMnF$_3$, RbCaF$_3$, and CsPbBr$_3$. \cite{hirota_neutron-_1995, ryan_observation_1986, songvilay_common_2019, hastings_central-peak_1978, cox_effect_1989, cox_effect_1989-1} Critical exponents for these systems range from 1 to 2 for the CP intensity and 0.5 – 0.84 for the correlation lengths, both much larger than what we observe. In fact, the temperature dependence of the CP in d$_6$-MAPI above $T_C$ is better described with a linear model. The lack of any divergent behavior is further indication that the cubic-tetragonal phase transition in d$_6$-MAPI is first-order. 

In summary, we have performed neutron inelastic scattering experiments of d$_6$-MAPI through the cubic-tetragonal phase transition and find that the CP model consistently provides a statistically better and physically more plausible description of the data. The \textit{R}-point scattering is dominated by a resolution-limited signal, and the temperature dependence does not behave as expected for dynamical domains. We cannot determine the physical origin of the CP signal, but it can be argued that phonon coupling is not maintained with the extremely short phonon lifetimes at the zone boundary, thus the defect-based origin is more likely. The data we obtained using the best energy resolution of 0.065 meV sets a lower bound on any dynamics of 36 ps at 340 K. We therefore conclude that the observed scattering is a manifestation of the CP phenomenon.

\begin{acknowledgments}
This work (neutron scattering) was supported by the Center for Hybrid Organic Inorganic Semiconductors for Energy (CHOISE), an Energy Frontier Research Center funded by the Office of Basic Energy Sciences, an office of science within the US Department of Energy (DOE). HIK acknowledges funding through the Department of Energy (DOE), Office of Basic Energy Sciences, Division of Materials Sciences and Engineering, under contract DE-AC02-76SF0051
\end{acknowledgments}

\bibliography{mt}
\bibliographystyle{apsrev}
\end{document}


\title{Supplemental Materials for: On the question of dynamic domains and critical scattering in cubic methylammonium lead triiodide}

\author{Nicholas J. Weadock}
 \affiliation{SSRL Materials Science Division, SLAC National Accelerator Laboratory, Menlo Park, CA 94025}
\author{Peter M. Gehring}%
\affiliation{NIST Center for Neutron Research, National Institute of Standards and Technology, Gaithersburg, MD 20899}%
\author{Aryeh Gold-Parker}
 \affiliation{Department of Chemistry, Stanford University, Stanford, CA 94305
}%
\author{Ian C. Smith}
\affiliation{Department of Chemistry, Stanford University, Stanford, CA 94305
}%
\author{Hemamala I. Karunadasa}
\affiliation{Department of Chemistry, Stanford University, Stanford, CA 94305
}%
\author{Michael F. Toney}
\affiliation{SSRL Materials Science Division, SLAC National Accelerator Laboratory, Menlo Park, CA 94025}
\date{\today}


\maketitle

\section{Determination of resolution functions}

The resolution function of neutron triple-axis spectrometers is well known to be an ellipse in four-dimensional ($\mathbf{Q}$,\,$\omega$) space determined entirely by the spectrometer parameters (collimators, monochromator and analyzer mosaicity, and spectrometer arm handedness, to name a few).\cite{shirane_neutron_2002} This ellipse is often expressed as a matrix $M(\mathbf{Q}$,\,$\omega)$, with elements proportional to the inverse linewidths of the Gaussian components of the resolution ellipse. In neutron triple-axis experiments, the measured signal is then a convolution of the resolution function $R(\omega - \omega_0,\mathbf{Q}-\mathbf{Q}_0)$ and sample scattering function $S(\mathbf{Q},\omega)$ as:
\begin{equation}
    I(\omega_0,\mathbf{Q}_0) = \int d\omega d\mathbf{Q}R(\omega - \omega_0,\mathbf{Q}-\mathbf{Q}_0)S(\mathbf{Q},\omega)
\end{equation}

Thus, extracting the sample scattering function requires knowing the magnitude and orientation of the resolution function at each measured ($\mathbf{Q}, \omega$). 

We characterized the resolution function two ways; experimentally and by calculating $M$ with the ResLibCal MATLAB package.\cite{farhi_ifit:_2014} The elastic incoherent energy width was determined by measuring a vanadium standard at $\mathbf{Q}$ = $\frac{1}{2}(1\, 3\, 3)$[reciprocal lattice units (rlu), 2.18$\mathrm{\AA^{-1}}$] in the d$_6$-MAPI coordinates. Additionally, the transverse and longitudinal Bragg widths ($\Delta\mathbf{Q_{\perp}}, \Delta\mathbf{Q_{\parallel}}$) were measured at either the $(2\,0\,0)$ or $(0\,1\,1)$ Bragg peaks of the cubic phase of d$_6$-MAPI. However, the resolution function changes with $\mathbf{Q}$ so these measurements are not enough to fully describe $M$. The SPINS and BT4 triple-axis spectrometer resolution matrices were calculated with ResLibCal using the Cooper-Nathans approximation (all Gaussian contributions to the resoultion function).\cite{shirane_neutron_2002} Figure~\ref{fig:Res011} below shows the results for SPINS at $E_f = 4$meV and $\mathbf{Q} = (0\,1\,1)$[rlu, $1.4 \mathrm{\AA^{-1}}$]. The measured elastic incoherent energy full-width at half-maximum (FWHM) is 0.19 meV, whereas the calculated width is 0.21 meV. Similarly, we measured a longitudinal elastic Bragg width of $\Delta\mathbf{Q_{\parallel}} = 0.016\mathrm{\AA}^{-1}$, and the calculated value is $0.015\mathrm{\AA}^{-1}$.

\begin{figure}
    \centering
    \includegraphics{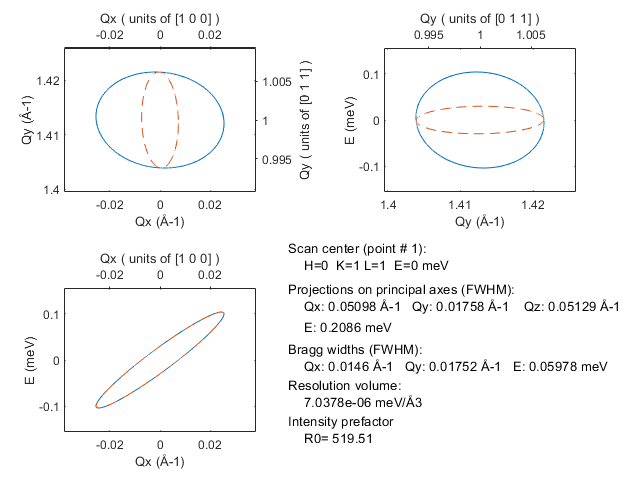}
    \caption{SPINS resolution function in several scattering planes calculated at $\mathbf{Q} = (0\,1\,1)$, $E_f = 4$meV. Cross-sections (Bragg widths) are shown as orange dashed lines, and projections as solid blue lines.}
    \label{fig:Res011}
\end{figure}

A deconvolution using the full resolution function is difficult and may not be necessary in certain situations. In the case of the central peak, the resolution limited component at $\hbar\omega=0$ means that $S(\textbf{Q}_i,\omega)$ at a given \textbf{Q} is effectively a delta function. Therefore, the wave-vector resolution function in \textbf{Q} is equivalent to the cross-sectional lineshape of the resolution ellipse at ($\textbf{Q}_i,\hbar\omega=0$), given as the Bragg width. In another extreme, if $S(\textbf{Q},\omega)$ is significantly wider than the resolution ellipse, the experiment will sample the entire width, projected onto the scattering plane. A constant-$\mathbf{Q}$ energy scan (E-scan) of a vanadium standard will return this projected resolution lineshape with an elastic incoherent FWHM.

\begin{figure}
    \centering
    \includegraphics{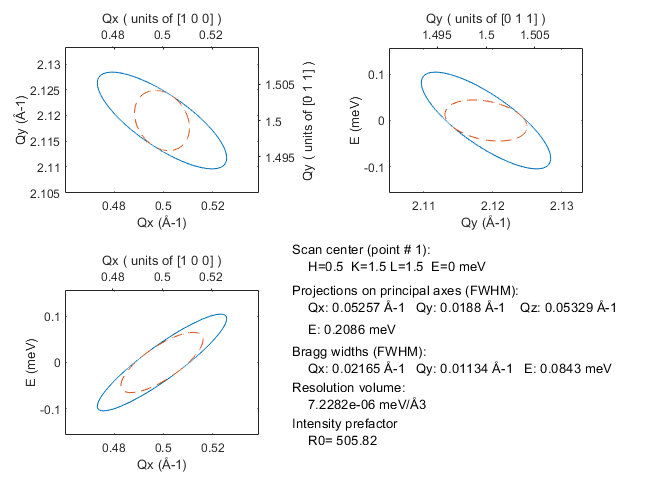}
    \caption{SPINS resolution function in several scattering planes calculated at the $d_6$-MAPI $R$-point, $\mathbf{Q} = \frac{1}{2}(1\,3\,3)$, $E_f = 4$meV. Cross-sections (Bragg widths) are shown as orange dashed lines, and projections as solid blue lines.}
    \label{fig:Res133}
\end{figure}

\begin{figure}
    \centering
    \includegraphics{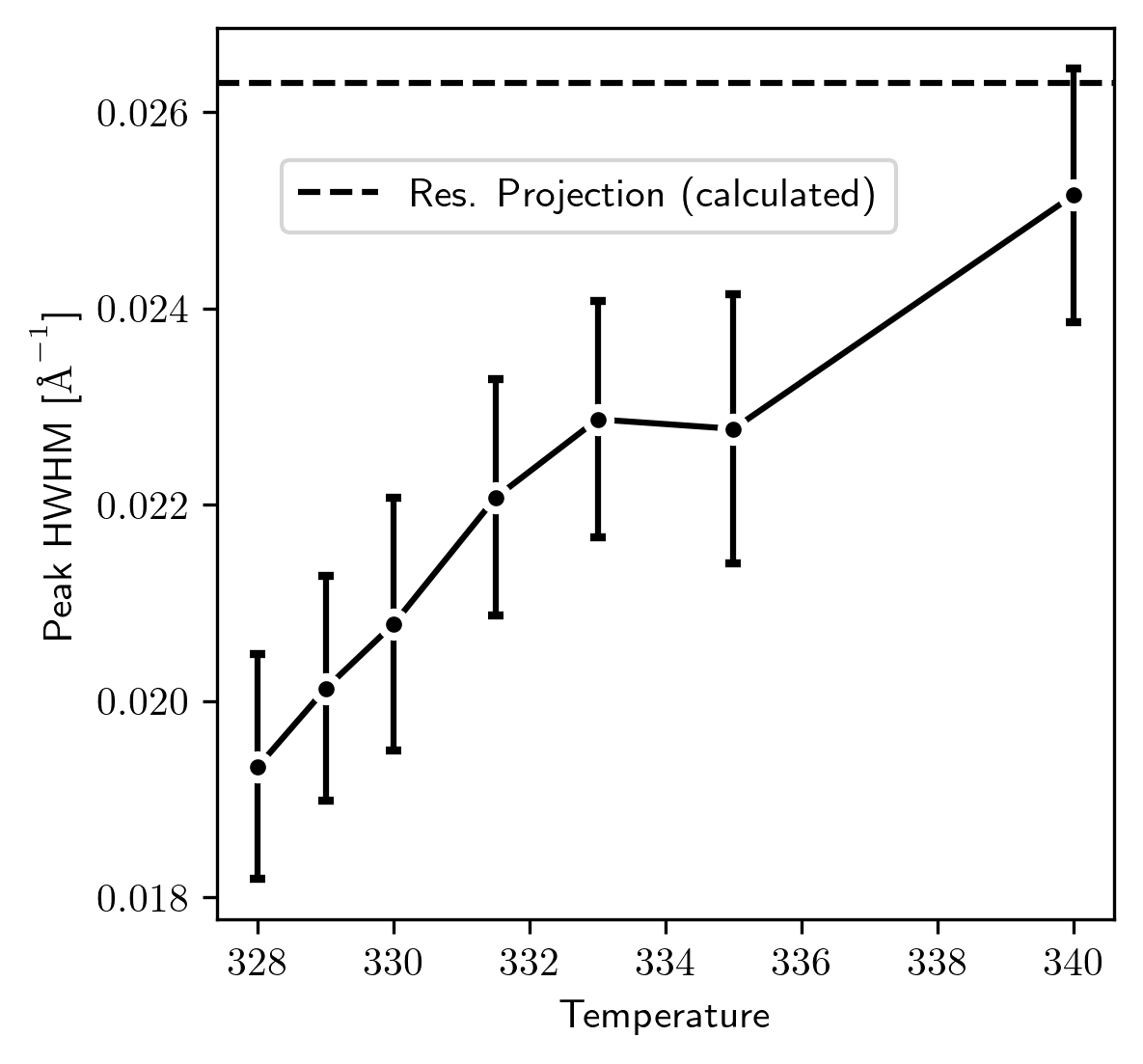}
    \caption{Resolution corrected $R$-point $\mathbf{Q}$-widths plotted as a function of temperature. The projected wave-vector resolution linewidth calculated for SPINS at $\mathbf{Q} = \frac{1}{2}(1\,3\,3)$, $E_f = 4$meV is indicated by the black dashed line.}
    \label{fig:rescompare}
\end{figure}

Complications arise when the widths of the scattering function change with experimental conditions and lie within the cross-sectional and projected widths of the resolution ellipse. The $\mathbf{Q}$-widths of the central peak are not resolution limited and have been demonstrated to decrease with temperature.\cite{shirane_q_1993} If our experimentally measured $\mathbf{Q}$-widths are less than the projected wave-vector linewidths, then the resolution linewidth in energy will be less than the elastic incoherent linewidth. We calculated $M$ at \textbf{Q} = $\frac{1}{2}(1\, 3\, 3)$[rlu], and the results are presented in Fig.~\ref{fig:Res133}. Our $\mathbf{Q}$-scans were along $\textbf{Q} = [1\, 0\, 0]$, and the corresponding projection $\mathbf{Q}$-width (FWHM) is $0.05\mathrm{\AA}^{-1}$. Figure~\ref{fig:rescompare} plots the measured linewidths as a function of temperature with the Bragg contribution removed. Below 340 K, the \textit{R}-point wave-vector linewidth is less than the projection $\mathbf{Q}$-width, and further decreases with temperature. 

\begin{figure}
    \centering
    \includegraphics[width=0.6\textwidth]{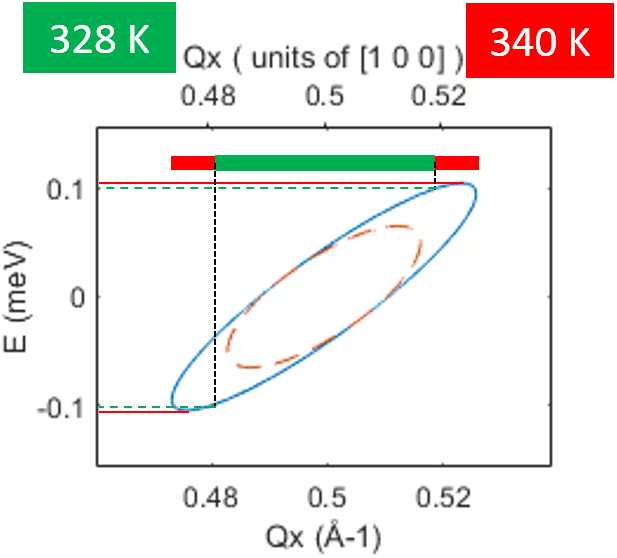}
    \caption{Schematic illustrating how much of the resolution ellipse is sampled (in energy), given a measured wave-vector linewidth of the scattering function $S(\mathbf{Q},\omega)$. At 340 K (red lines), the measured linewidth is greater than the projected resolution ellipse, thus the energy resolution is equal to the projection cross-section. The linewidth at 328 K (black dashed lines) is less than the projected width, and the resulting energy linewidth is indicated by the green dashed lines. Due to the orientation of the resolution ellipse, only a negligible narrowing in energy resolution results.}
    \label{fig:schematic}
\end{figure}

We use the schematic in Fig.~\ref{fig:schematic} to demonstrate the effect of the narrow $\mathbf{Q}$-width on the resolution function in energy. At 340 K, the full projection width is sampled, shown by the solid red lines originating from the blue projection ellipse. When the temperature is reduced to 328 K, the narrowed $\mathbf{Q}$-width intersects the resolution ellipse as indicated by the black dashed lines. The corresponding energy width is set by the dashed green lines. Despite the large difference in the measured $\mathbf{Q}$-width and the projection $\mathbf{Q}$-width, the orientation of the resolution ellipse in the scattering plane is such that the projected energy width is only slightly reduced.

Based on the above analysis, we choose to fit the E-scans and $\mathbf{Q}$-scans using the experimentally determined 1-D elastic incoherent energy resolution and Bragg cross-sectional wave-vector resolution. One-dimensional peak fitting was performed using a non-linear least squares fitting algorithm in DAVE.\cite{azuah_dave_2009} A full deconvolution could be performed, however additional uncertainty would be introduced by the fact that the calculated $M$ does not match the experimentally determined values.

\pagebreak

\section{Linewidth comparison for Elastic $\mathbf{Q}$ scans}

\begin{figure}[!h]
    \centering
    \includegraphics{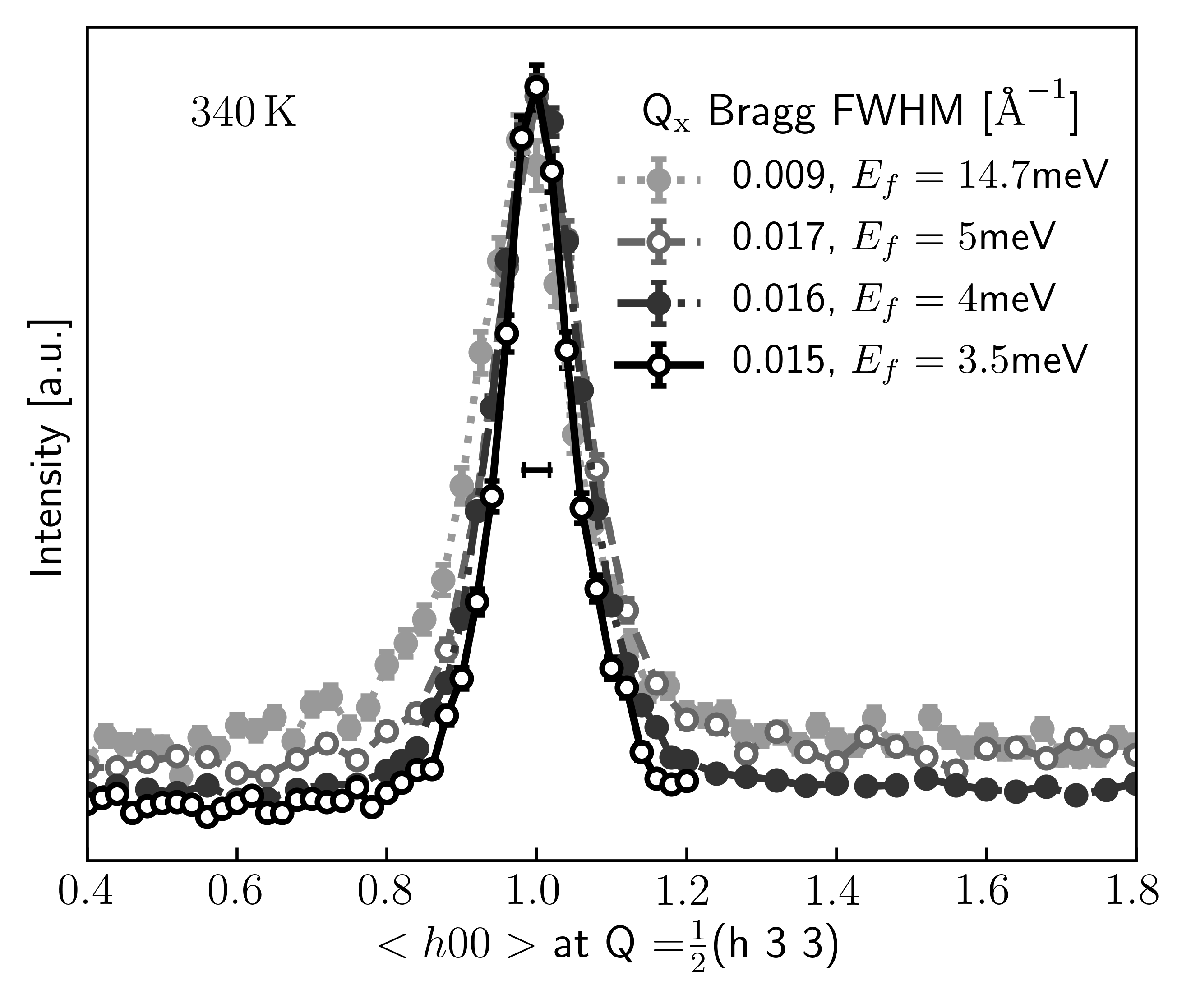}
    \caption{Elastic $\mathbf{Q}$-scans measured at 340 K with four different resolution cross-sections (FWHM reported in legend) obtained on SPINS ($E_f = 3.5,\,4,\,5$ meV) and BT4 ($E_f = 14.7$ meV). The horizontal bar represents the largest FWHM from all four resolution configurations.}
    \label{fig:QscanCompare}
\end{figure}

The $\mathbf{Q}$-scan cross-sectional linewidths (Bragg widths) for BT4 were determined by measuring the (200) Bragg peak (\textbf{Q} = 2.0$\mathrm{\AA}^{-1}$) of d$_6$-MAPI at 350 K, whereas for SPINS the Bragg widths were measured at the (011) peak (\textbf{Q} = 1.4$\mathrm{\AA}^{-1}$) at 340 K. Elastic $\mathbf{Q}$-scans at 340 K for all four experimental configurations are plotted in Fig.~\ref{fig:QscanCompare}, along with the largest resolution cross-sectional linewidth. It is apparent that the scattering is not resolution limited.
\newpage
\section{Central peak and dynamic-domain model comparison}

\begin{figure}
    \centering
    \includegraphics{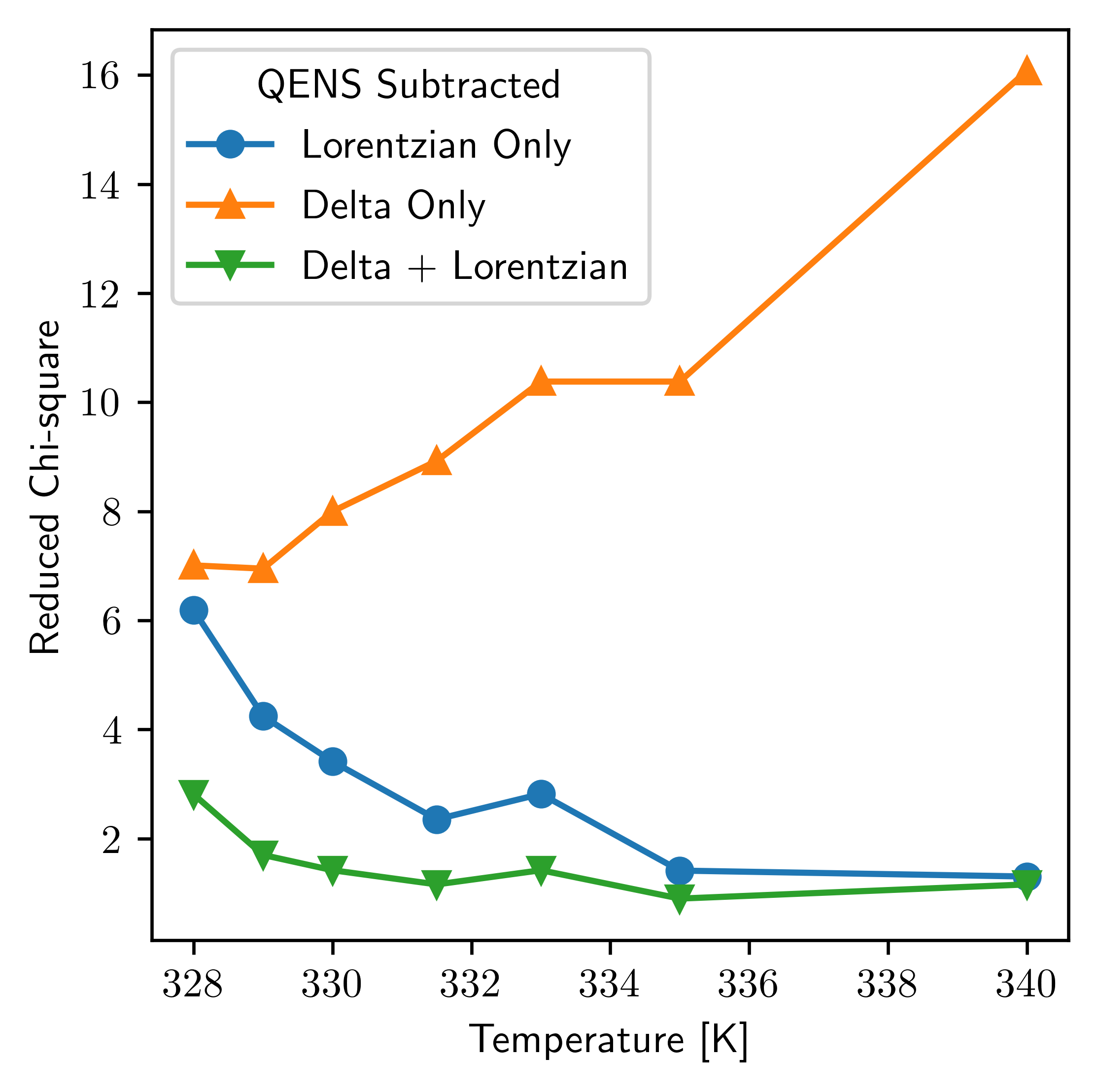}
    \caption{Reduced-$\chi^2$ statistic as a function of temperature for three different fit models; the resolution function (orange triangle), a resolution-broadened Lorentzian representing dynamic domains (blue circle), and a resolution-limited central peak plus Lorentzian (green upside-down triangle).}
    \label{fig:chisq}
\end{figure}

The validity of central peak (CP) and dynamic tetragonal domain models are also assessed by comparing the reduced-$\chi^2$ goodness-of-fit statistic for each model as a function of temperature. As stated in the Main text, our CP model includes a delta function plus a Lorentzian, convolved with the resolution function. A model describing dynamic tetragonal domains consists of a single Lorentzian, convolved with the resolution function. For comparison, we also fit the measured data to just the resolution function itself. In all cases, the quasielastic scattering (QES) contribution is subtracted before fitting, as explained in the main text. Figure~\ref{fig:chisq} plots the reduced-$\chi^2$ goodness-of-fit statistic for each model. At 340K, the CP and dynamic-domain models fit the data equally well. As the transition temperature ($327.5~\pm~0.5$K) is approached, however, the dynamic-domain model gives results that are much poorer fits to the data. The pure resolution function is the poorest description, but improves with decreasing temperature. 

\begin{figure}
    \centering
    \includegraphics{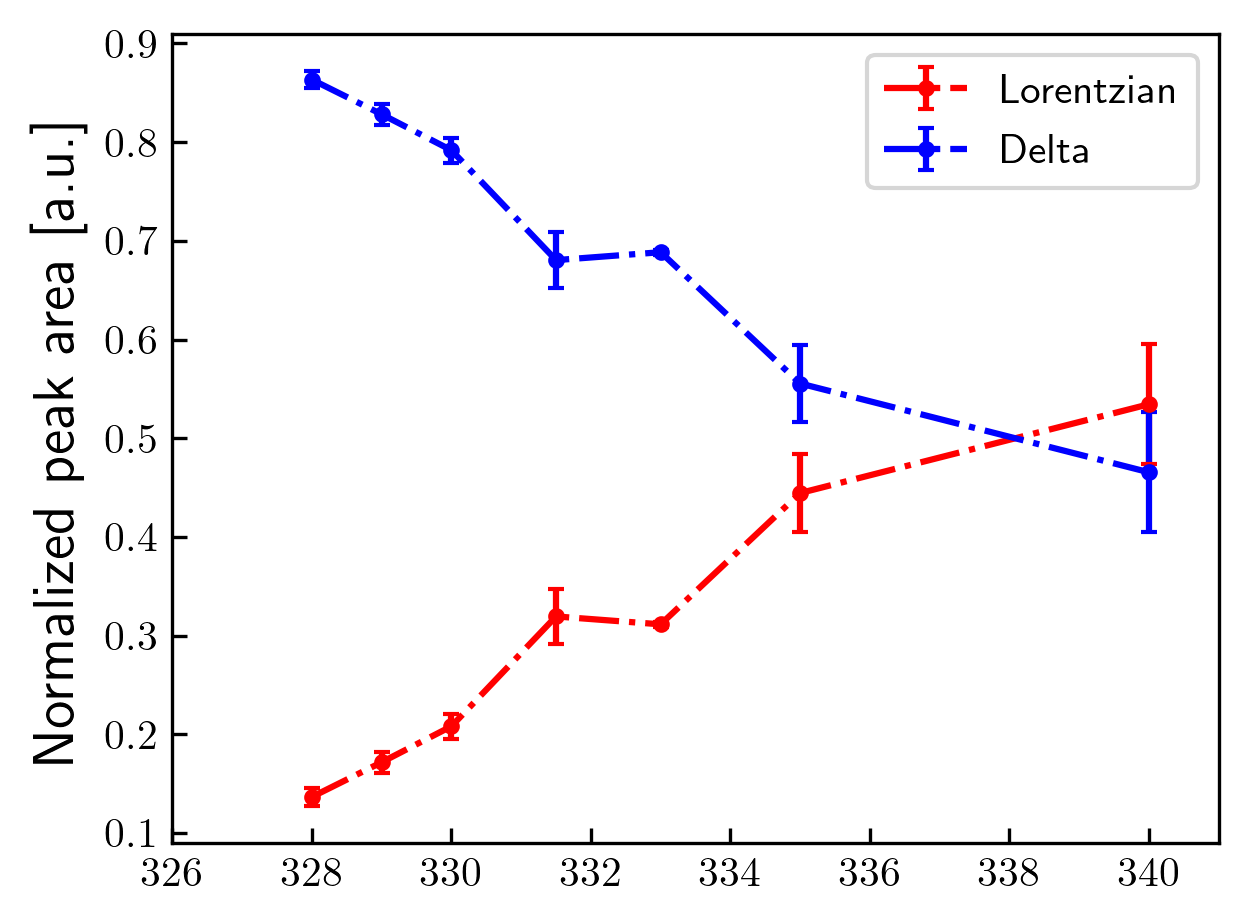}
    \caption{Temperature dependence of the relative area fractions of the delta function and Lorentzian components }
    \label{fig:areafrac}
\end{figure}

\begin{figure}
    \centering
    \includegraphics{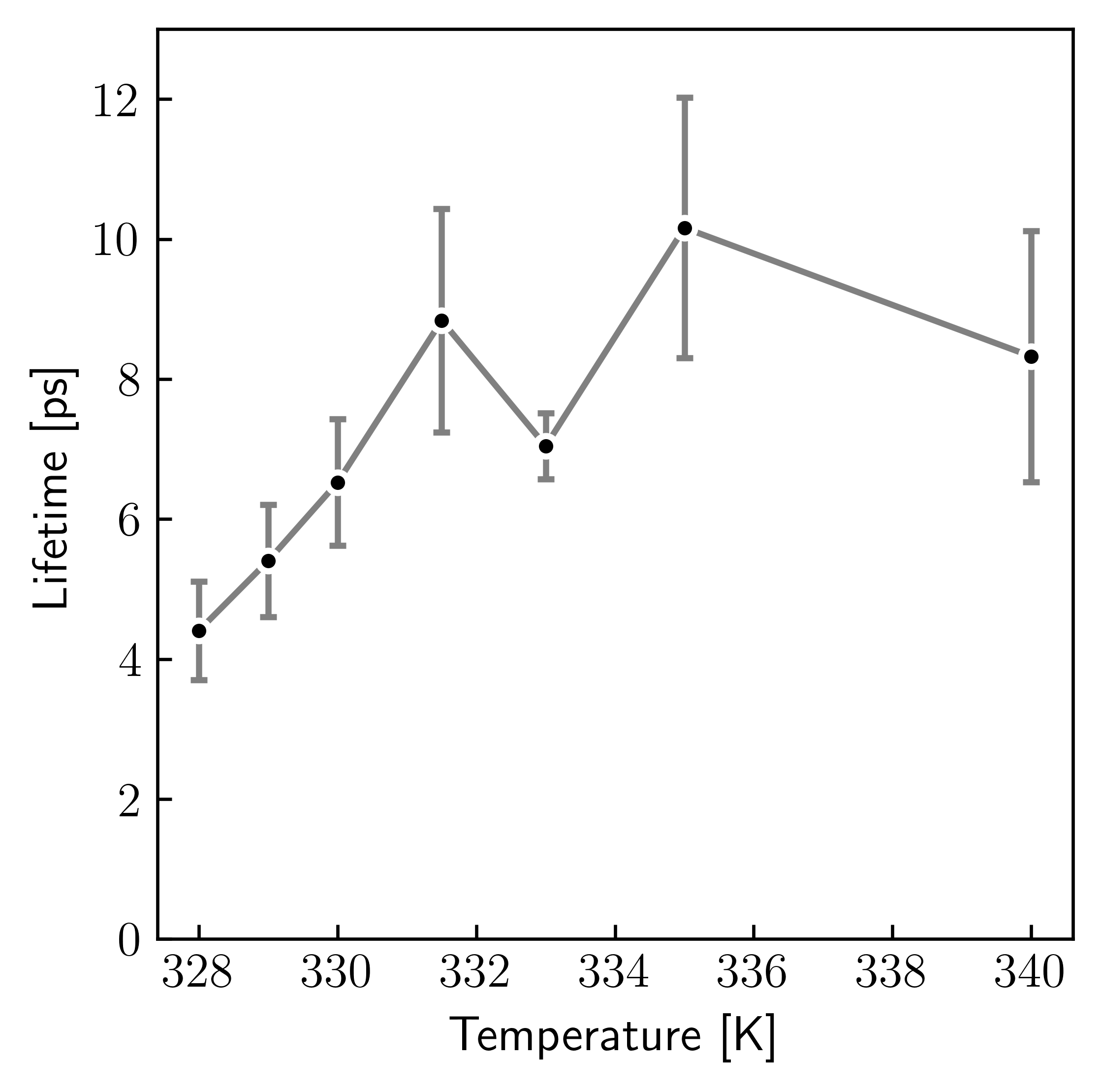}
    \caption{Lifetimes obtained from the Lorentzian component of the CP model, evaluated as $\tau = \hbar/\mathrm{HWHM}$.}
    \label{fig:CPlife}
\end{figure}

The relative area fractions of the delta function and Lorentzian components within the CP model are plotted in Fig.~\ref{fig:areafrac} as a function of temperature. As the temperature decreases towards $T_C$, the resolution-limited delta function area increases. The linewidth of the remaining Lorentzian component is converted to a lifetime as $\tau = \hbar/\mathrm{HWHM}$, and plotted in Fig.~\ref{fig:CPlife}. We see that the lifetime does not exhibit a strong temperature dependence, and in fact decreases as the transition temperature is approached. This behavior is not expected for dynamic domains.

\newpage

\section{Estimating a lower bound on dynamic domain lifetime}

Our analysis in the main text indicates that the observed critical scattering is a manifestation of the central peak phenomenon. However, if a Lorentzian component describing dynamic domains is significantly narrower than the resolution function, the signal will be masked by the broader resolution function.\cite{stirling_critical_1996} We seek to establish a lower bound on the lifetime of dynamic domains by determining the broadest Lorentzian which would still return an apparent resolution-limited linewidth when convolved with the measured elastic incoherent resolution function. 

\begin{figure}
    \centering
    \includegraphics{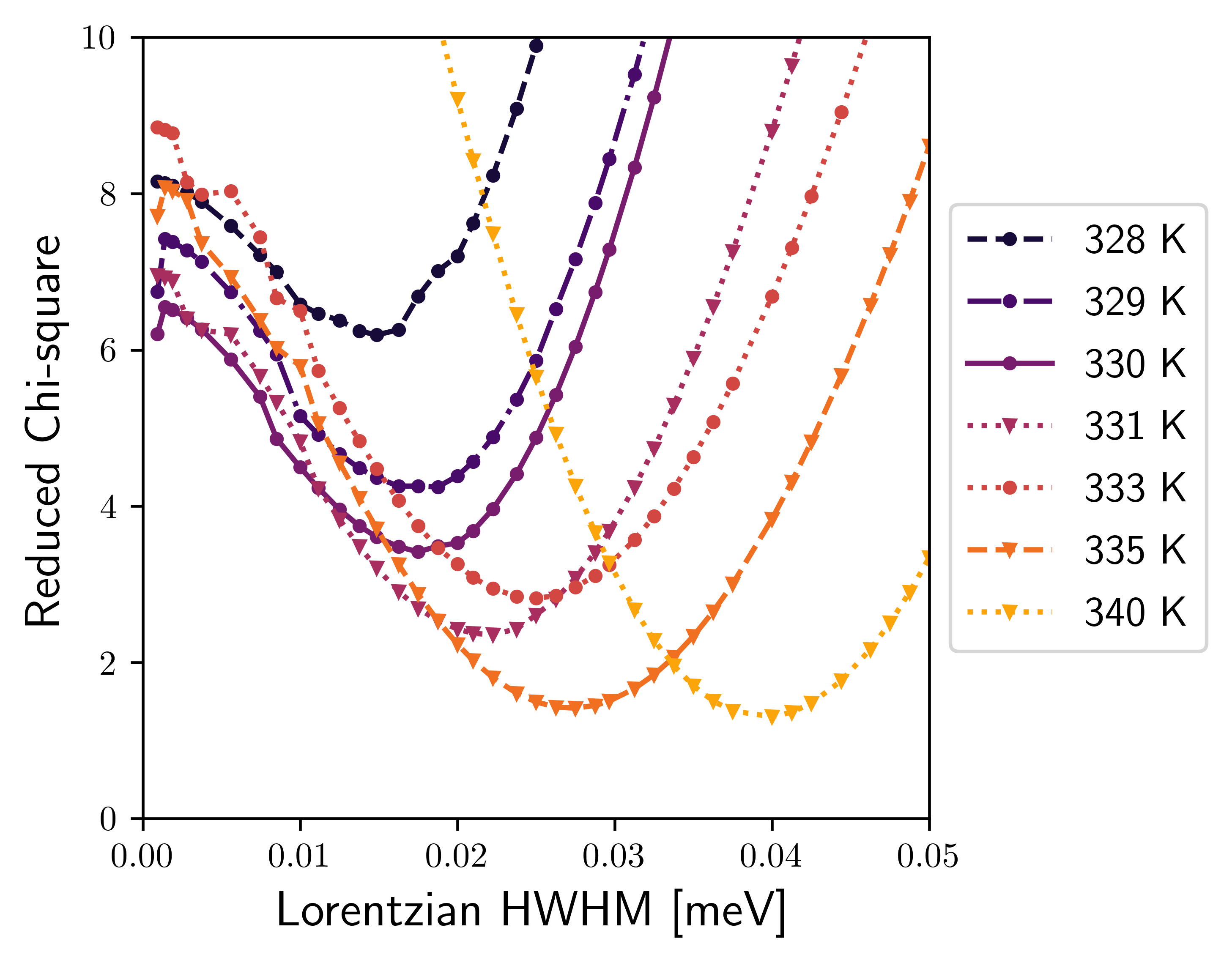}
    \caption{Reduced-$\chi^2$ goodness of fit statistic obtained from least squares fitting of resolution-broadened fixed-width Lorentzians to \textit{R}-point E-scans with resolution HWHM = 0.095 meV.}
    \label{fig:chiwidth}
\end{figure}

To do this, QES subtracted \textit{R}-point E-scans at each temperature were fit with a non-linear least-squares routine (\texttt{lmfit} package)\cite{matt_newville_2019_3588521} to a series of resolution broadened Lorentzians with fixed HWHM and variable area. The resulting reduced-$\chi^2$ statistic as a function of Lorentzian HWHM is plotted in Fig.~\ref{fig:chiwidth}. As the temperature decreases towards the ciritcal temperature, the best-fitting linewidth also decreases, however the overall goodness of fit is worse (increased reduced-$\chi^2$). 

We assume that the data points are normally distributed with a probability distribution given by:
\begin{equation}
    f(x; \theta, \sigma) = \frac{1}{\sigma\sqrt{2\pi}}\mathrm{e}^{-\frac{1}{2}\left(\frac{x - \theta}{\sigma}\right)^2}
\end{equation}
where \textit{x} is the measured value, and $\theta$, $\sigma$ are the expected value and variance according to the proposed model. We seek the likelihood that the observed data is described by a resolution-broadened Lorentzian with fixed HWHM,
\begin{equation}
    L(\theta, \sigma; x_i) =
    \frac{1}{\sigma_i\sqrt{2\pi}}\mathrm{e}^{-\frac{1}{2}\sum\limits_i\left(\frac{x_i - \theta_i}{\sigma_i}\right)^2}
\end{equation}

Using the definition of the $\chi^2$ statistic, we obtain:
\begin{equation}
    L(\chi^2; x_i) \propto \exp(-\chi^2 /2)
\end{equation}

\begin{figure}
    \centering
    \includegraphics{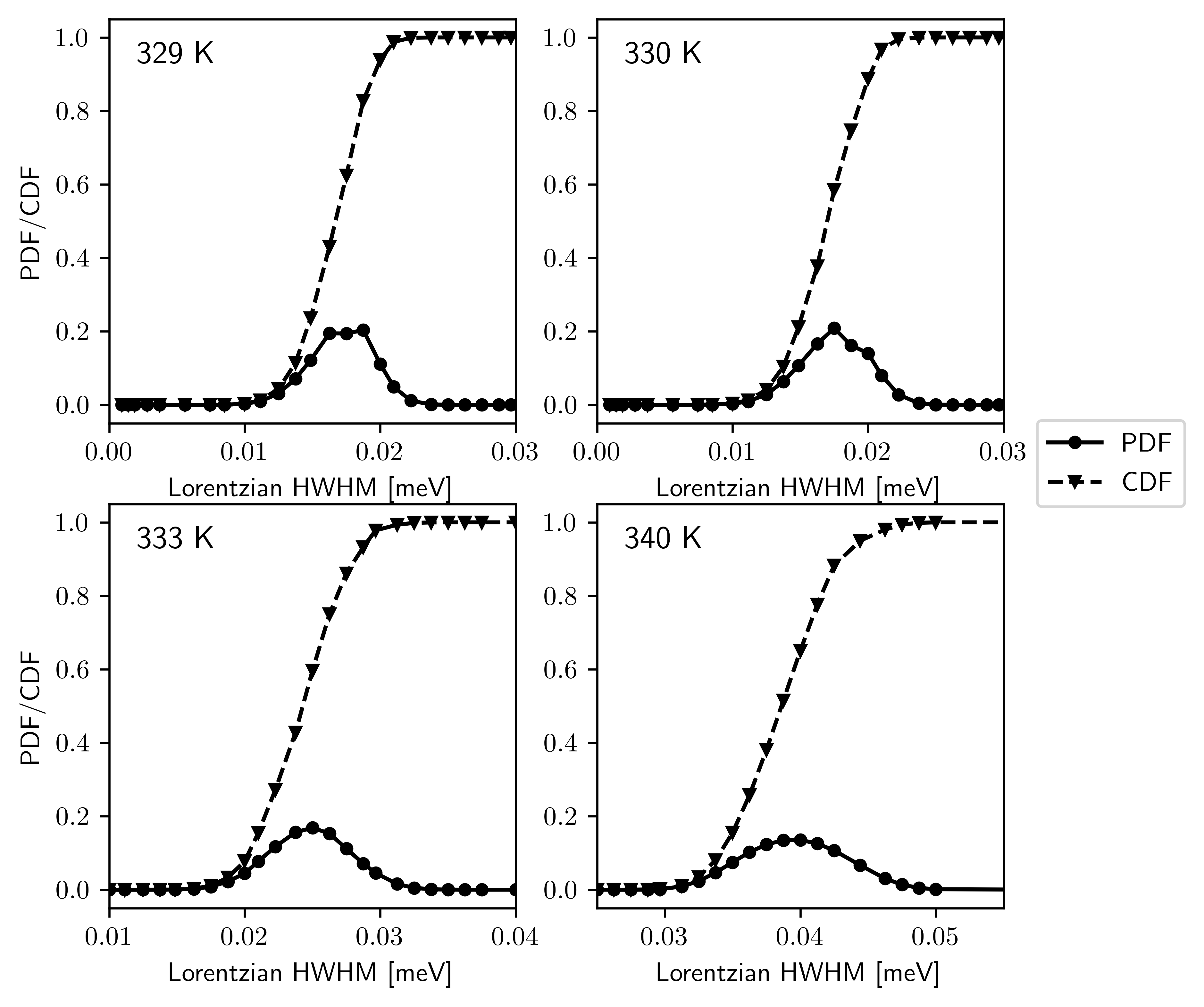}
    \caption{Select posterior distribution functions (and corresponding cumulative distribution functions) derived from the log likelihood that the observed data is described by a resolution-broadened Lorentzian with fixed HWHM.}
    \label{fig:pdf}
\end{figure}

We use the log of this likelihood function, together with a uniform prior, to generate a posterior distribution function (PDF) according to Bayes' theorem:
\begin{equation}
    P(x;\chi^2) = C(\exp-\chi^2 /2) 
\end{equation}
where C is a normalization constant determined from the full range of sampled HWHM. Representative PDFs at four temperatures, and their corresponding cumulative distribution functions (CDF), are plotted in Fig.~\ref{fig:pdf}.

\begin{figure}
    \centering
    \includegraphics{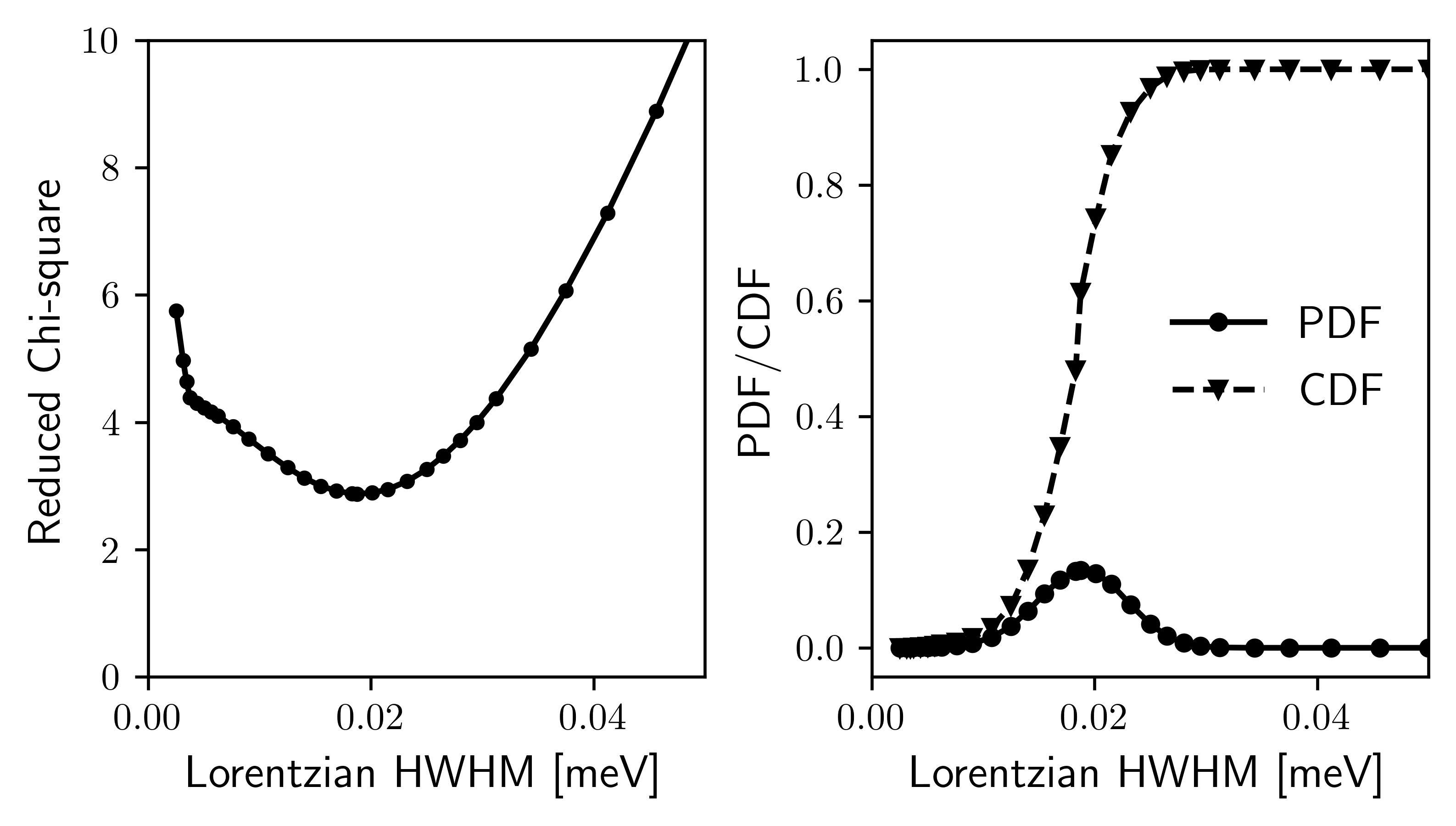}
    \caption{Left: Reduced-$\chi^2$ goodness-of-fit statistics for least squares fitting of resolution-broadened fixed-width Lorentzians to the 340 K \textit{R}-point constant-$\mathbf{Q}$ scan with resolution HWHM = 0.065 meV. Right: Posterior and cumulative density distribution functions derived from log likelihood that the observed scattering lineshape is described by a resolution-broadened Lorentzian with fixed HWHM.}
    \label{fig:Ef35minlft}
\end{figure}

The obtained CDFs are used to construct a 95\% highest posterior density interval (HDPI) within the regions illustrated in Fig.~\ref{fig:pdf}. The HWHM at the center and ends of this interval are converted to a lifetime according to $\tau$ = $\hbar$/HWHM \cite{maradudin_scattering_1962} and reported in Fig. 3a) of the main text. The same analysis was performed for the spectra obtained with a resolution HWHM = 0.065 meV, and results are presented in Fig.~\ref{fig:Ef35minlft}.

\newpage

\section{Lattice Parameter measurements}

\begin{figure}
    \centering
    \includegraphics{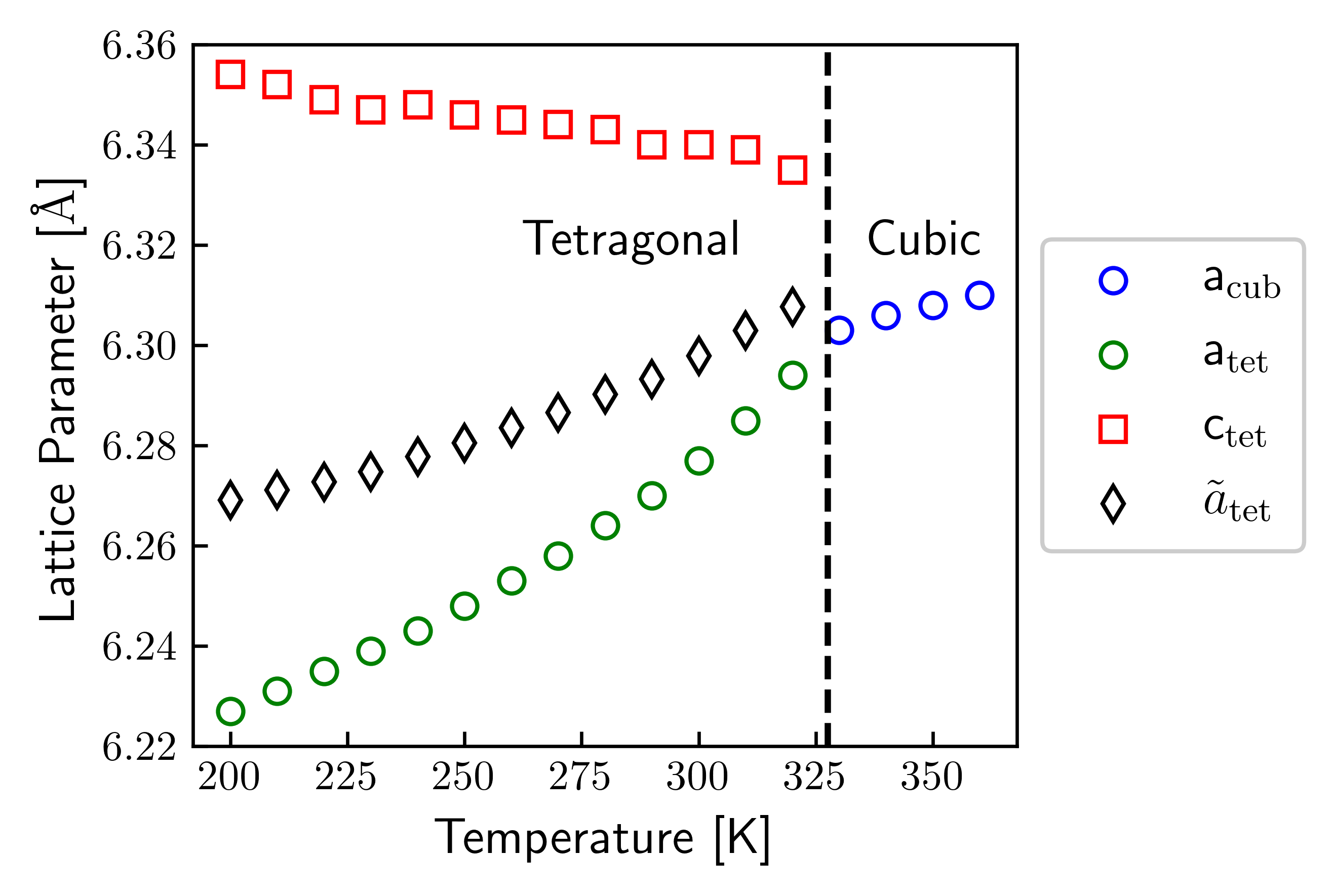}
    \caption{Lattice parameter of d$_6$-MAPI as measured from the (2 0 0) reflection on SPINS.}
    \label{fig:Latt}
\end{figure}

The lattice parameter of the cubic and tetragonal phases were measured at the (2 0 0) reflection on SPINS ($E_f = 5$ meV) down to 200 K, and the results are plotted in Fig.~\ref{fig:Latt}. The pseudocubic lattice parameter, calculated as $\tilde a_{\mathrm{tet}} = (a_{\mathrm{tet}}*c_{\mathrm{tet}})^{\frac{1}{3}}$ is also plotted in the tetragonal phase. These results are consistent with other reported lattice parameters.\cite{kawamura_structural_2002, whitfield_structures_2016}

A discontinuity in the pseudocubic lattice parameter across the transition, and the fact that the c/a ratio does not go to 1 indicates a first-order transition. The c/a ratio from lattice parameters reported by Kawamura, et al., and Whitfield, et al., do not go to 1 either.\cite{kawamura_structural_2002, whitfield_structures_2016} However, Kawamura et al., report no discontinuity in the pseudocubic lattice parameter while there appears to be a jump in Figure 6c) of \cite{whitfield_structures_2016}. Both groups suggest the phase transition could be close to tricritical, although Kawamura et al., concede that heat capacity measurements indicate a first-order transition.

\newpage
\bibliography{supplemental}
\bibliographystyle{apsrev}